# Review of Automaton Learning Algorithms with Polynomial Complexity - Completely Solved Examples

Dr. Farah Haneef
Computer Science Department
Quaid-e-Azam University, Islamabad, Pakistan

## 1 Introduction

Automaton learning is a domain in which the target system aka System Under Learning (SUL) is inferred by the automaton learning algorithm in the form of an automaton, by synthesizing a finite number of inputs and their corresponding outputs. Automaton learning makes use of a Minimally Adequate Teacher (MAT). The learner learns the SUL by posing membership queries to the MAT.

In the early stages, theoretical automaton learning has successful real-world applications. It goes beyond formal verification and allows to infer behavioral model of black-box systems. The improvements in tool support and raising competition focuses on the practicality of automaton learning and encouraging Learning-based Testing (LBT) techniques, which shows the growing interest in the field of research.

Learning-based testing [10] is an emerging paradigm of software testing. It is a heuristic iterative approach which is useful to automate specification based black box testing [9][7]. The LBT framework consists of a system under test (SUT), a formal specification for SUT and a learned model of SUT. A learning base testing algorithm works by executing test case inputs on the SUT. Most of the learning algorithms learn in the limit to yield a minimal approximation of the target DFA. This concept of learning in the limit for target DFA was first introduced by E.M.Gold in 1967 [5] . In his paper, he showed that with the help of some inference or learning algorithm, a regular language corresponding to some target DFA can be guessed by a finite number of wrong hypothesis. With respect to learning type, there are three kinds of learning algorithms: *complete* learning algorithms, *incremental* learning algorithms and *sequential* learning algorithms.

In complete learning algorithms, initially, the system under learning (SUL) is completely learned by giving different inputs and receiving their corresponding outputs. After complete learning of the system, a hypothesis *H* is generated [1]. While in incremental learning algorithms, system is learned in incremental form [13]. Initially, a small part (increment) of the system is learned by giving the inputs and receiving corresponding outputs after that, these input/output pairs are synthesized into a hypothesis $H_i$ by the learning algorithm in an incremental fashion. The next step it learns a new increment (small part of the system) and rebuilds the hypothesis $H_{i+1}$ which contains the previously and newly learned information. This process of learning can potentially continue till the complete learning of the system. On the other hand, sequential learning algorithm similarly works as an incremental learning algorithm except that it does not reuse the previously learned information from the hypothesis $H_{s-1}$ to construct the new hypothesis $H_s$ [8].

In automaton learning, there is a *learner* (learning algorithm) and an *adequate teacher* [2]. The Learner is a learning algorithm which learns the regular set from queries and counter-examples. The Learner asks queries to the adequate teacher. The adequate teacher answers the questions from the Learner, about the unknown regular set. It answers two types of questions: First type is a *membership query*, consisting of a string $t \epsilon \Sigma^*$. The adequate teacher answers as *yes* or *no* depending on whether string $t$ is a member of the unknown set or not. The second type of question is a *conjecture*, consisting of a description of a regular set $S$; the answer is *yes* if $S$ is equal to the unknown language and is a string $t$ in the symmetric difference of $S$ and the unknown language otherwise. In the second case, the string $t$ is called a counter-example because it serves to show that the conjectured set $S$ is incorrect.

The concept of the adequate teacher was first introduced by Dana Angluin, in ID algorithm [1]. After that this concept was used by other researchers. As they found that if there is an adequate teacher, the complexity of automaton learning is polynomial whereas, in the absence of the adequate teacher automaton learning is an NP-hard problem [10].

## 2 Automaton Learning Algorithms

We have briefly studied some automaton learning algorithms with the help of examples, which are specifically relevant to table data structure and tree data structure. The learning algorithms having table data structure are L* proposed by Dana Angluin in 1987 [2], ID proposed by Dana Angluin in 1981 [1], DLIQ and BDLIQ proposed by Farah Haneef and Muddassar Azam Sindhu in 2022 and 2023 respectively in [15, 16], IID proposed by R. Parekh, C. Nichitiu and V. Honavar in 1998 [13], IDS proposed by K. Meinke and Muddassar Azam Sindhu in 2010 [11], IDLIQ proposed by Farah Haneef and Muddassar Azam Sindhu in [17] and IKL proposed by K. Meinke and Muddassar Azam Sindhu in 2011 [12]. The learning algorithms having tree data structure are RPNI proposed by Oncina and Garcia in 1992 [3] and RPNII proposed by Pierre Dupont in 1996 [4].

### 2.1 L* Algorithm

The L* is a complete learning algorithm proposed by Dana Angluin in 1987 [2]. It learns a regular set by asking membership and equivalence queries. A membership query tells whether a string $\alpha$ is a member of the language of DFA $A$ or not, $\alpha \epsilon L(A)$? An equivalence query determines whether a hypothesis DFA is a correct representation of regular set or not i.e. $L(H) = L(A)$? L* asks membership queries and organizes this information in form of a table consisting of a tuple (S, E, T) which is called Observation Table *OT*. Where $s_1, s_2, \ldots, s_n$ are row labels belonging to *S* and $e_1, e_2, \ldots, e_n$ are column labels belonging to *E*. Whereas *T* is a transition function $((S \cup S \cdot \Sigma) \times E)$. A simple observation table (*OT*) is shown in Table 2.1.1. This table has two parts. Upper part consists of set $S = \{s_1, s_2\}$ and lower part consists of concatenation of *S* and *Σ* having $s_1. a_1$ and $s_2. a_2$ elements, where $s_1, s_2 \in S$ and $a_1, a_2 \in \Sigma$ and *Σ* is a finite set of alphabets.

Function *row(s)* is a finite function which represents the tuple of entries in the observation table *OT* corresponding to row labeled as *s*.

| T | $e_1$ | $e_2$ |
|---|---|---|
| $s_1$ | | |
| $s_2$ | | |
| $s_1.a_1$ | | |
| $s_2.a_2$ | | |

Table 2.1.1: Observation Table (*OT*)

This table should meet two basic properties before asking equivalence queries to make conjecture. These properties are *closure* and *consistency*. The *OT* is called closed if and only if, for each string *t* in lower part of the table i.e. *S.Σ*, there exist an *s* in upper part of the table i.e. *S* such that *row(t)* = *row(s)*. If *OT* is not closed then rows of the observation table are extended as *S* with prefixes of *S*. For *OT* to be consistent, it is necessary that if any two rows of upper part of *OT* are same as *row(s1)* = *row(s2)* then for all $a \in \Sigma$, *row* $(s_1. a)$ = *row* $(s_2. a)$. If the observation table is closed but not consistent then column of observation table is extended with a symbol *a* where $a \in \Sigma$.

When *OT* is closed and consistent, a conjecture can be constructed. Distinct *rows(S)* show the different states (states described by concatenation of elements of set *E*) and column *T* represents the strings *t* which are used to show the transitions from one state to another. The initial state is the *row(λ)* and the final states are the entries in the table under the column *λ* with value = 1. We start from the initial state $q_0$ and check all transitions of input alphabet, existing in Column *T* as *δ(row(s), a)* = *row (s. a)*. After reading all input symbols from each state, a conjecture is represented in the form of a table. Rows represent the states as $q_0, q_1, \ldots, q_n$ and columns represent input symbols as 0, 1.

If this conjecture *H* is language equivalent to the target DFA, *A*, *L(H)* = *L(A)* then the adequate teacher answers the equivalence query as Yes otherwise, it answer as No and gives the counter example either from *L(H)* - *L(A)* or *L(A)* - *L(H)*, which is then accommodated in *OT* in the form of extension of rows. This process continues until *OT* becomes consistent and closed.

## Example

An example run of the L* algorithm is given below:
Unknown Regular Set: *U* = Even number of 0's except the empty string.
Fixed known Finite Alphabets: $\Sigma = \{0, 1\}$
The initial observation table is shown in Table 2.1.2. It shows that $S = \{\lambda\}$ and $E = \{\lambda\}$. The corresponding table entries are called as: *λ. λ = λ* which is not an accepting string so value = 0 is inserted in the corresponding cell. $0.\lambda = 0$ which shows odd number of 0's which are not accepted so value $= 0$, and $1.\lambda = 1$ which shows zero number of 0's therefore the corresponding cell value $= 0$.

| $T_1$ | $\lambda$ |
|---|---|
| $\lambda$ | 0 |
| 0 | 0 |
| 1 | 0 |

Table 2.1.2: Initial Observation Table

As in upper part of $\lambda$ column there is no cell having value 1 therefore this table shows that there is no final state in the initial hypothesis DFA. Initial *OT* is closed as *row* (0) = *row* ($\lambda$) and *row* (1) = *row* ($\lambda$). This *OT* is also consistent as no two rows in *S* are same. Now L* makes a conjecture shows that if we read 0 or 1 from *state0* then it will remain to the same state as itself. Conjecture $H_1$ and respective automaton is shown in Fig.2.1.1: (a) and (b)

| $\delta$ | 0 | 1 |
|---|---|---|
| $q_0$ | $q_0$ | $q_0$ |

Fig.2.1.1(a): Conjecture $H_1$

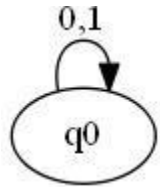

Fig.2.1.1(b): Automaton $H_1$

Above conjecture $H_1$ shows that there is no final state in it which makes its language inequivalent to the target DFA, *A*. Let $H_1$ gives counter-example as 00 which is an accepting string on target DFA but above automaton does not accept it. Therefore, we extend the rows of initial *OT* $T_1$ as 0, 00 in *S* and 01, 001, 000 in *S.Σ*. The extended observation table $T_2$ is shown below as Table 2.1.3.

| $T_2$ | $\lambda$ |
|---|---|
| $\lambda$ | 0 |
| 0 | 0 |
| 00 | 1 |
| 1 | 0 |
| 01 | 0 |
| 001 | 1 |
| 000 | 0 |

Table 2.1.3: Observation Table $T_2$

Table 2.1.3 shows that $\lambda$. $\lambda = \lambda$ so value = 0, 0.$\lambda$ = 0 so value = 0, 00.$\lambda$ = 00 which has even number of 0's so it is an accepting string and the value against it will be 1. Similarly, 1.$\lambda$ = 1 so value = 0, 01. $\lambda$ = 01 so value = 0, 000. $\lambda$ = 000 so value = 0, 001.$\lambda$ = 001 having two 0's so this string is also accepted and value against it will be 1. $T_2$ has $S = \{\lambda, 0, 00\}$ and $E = \{\lambda\}$

Observation table $T_2$ is closed as for every successor state in $S.\Sigma$, there is a row in $S$ as $row(s) = row(t)$ not consistent as $row(\lambda) = row(0)$ but $row(0)$ is not equivalent to $row(00)$. Therefore, we have to extend set $E$ with 0. Observation table $T_3$ is shown below as Table 4 having $S = \{\lambda, 0, 00\}$ and $E = \{\lambda, 0\}$

| $T_3$ | $\lambda$ | 0 |
|---|---|---|
| $\lambda$ | 0 | 0 |
| 0 | 0 | 1 |
| 00 | 1 | 0 |
| 1 | 0 | 0 |
| 01 | 0 | 1 |
| 001 | 1 | 0 |
| 000 | 0 | 1 |

Table 2.1.4: Observation Table $T_3$

In Table $T_3$ we have called the column 0 as: $\lambda.0 = 0$ as number of 0's is 1 so value=0, 0.0 = 00 as number of zeros are two so corresponding value = 1, 00.0 = 000 here number of zeros are three so value = 0 and similarly other values of column 0's are calculated for the lower part of the table.

$T_3$ is closed as well as consistent as $row(\lambda) = row(00)$ and $row(1) = row(001)$ and $row(0) = row(000)$. So, L* makes a conjecture $H_2$ which is shown in Fig.2.1.2(a). Rows against column $\lambda$ and 0 shows the states of automaton as 00 = $q_0$, 01 = $q_2$, 10 = $q_2$. whereas transitions are as follows.

If we read strings from column $T_3$ as 0 from the *state*00 then we reach the *state*01. If we read 1 from *state*00 will reach the *state*00 (self-loop). If we read the string 0 from the *state*01 we reach the *state*10. If we read the string 1 from the *state*01 we reach the *state*01 (self-loop). Similarly, if we read the string 0 from the *state*10, we reach the *state*01 and if we read the string 1 from the *state*10, we reach the *state*10.

| $\delta$ | 0 | 1 |
|---|---|---|
| $q_0$ | $q_1$ | $q_0$ |
| $q_1$ | $q_2$ | $q_1$ |
| $q_2$ | $q_1$ | $q_2$ |

Fig.2.1.2(a): Conjecture $H_2$

As $H_2$ accepts all the strings having even number of 0's except the null string so the adequate teacher answers as Yes. L* terminates and gives a minimal behaviorally equivalent automaton $H_2$ to the target automaton $A$ shown in Fig.2.1.2(b).

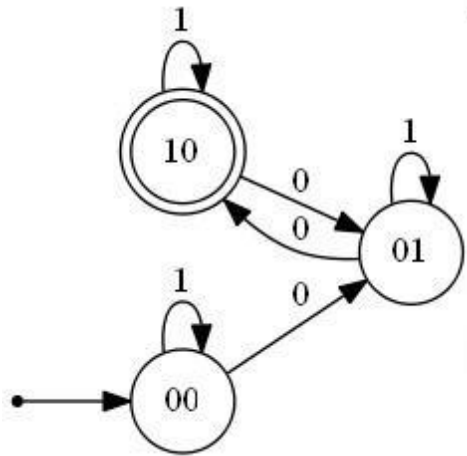

Fig.2.1.2(b): Automaton $H_2$

## 2.2 ID Algorithm

The ID algorithm is a complete learning algorithm proposed by Dana Angluin in 1981 [1]. It asks membership queries from the adequate teacher to learn the regular set. It uses the concept of live states and dead state. A state $q_1 \in Q$ is called a live state if there exists a string $\sigma_1, \sigma_2, \ldots, \sigma_n \in \Sigma^*$ such that $\delta^*(q_0, \sigma_1, \sigma_2, \ldots, \sigma_n) = q_i$ and $q_i \in F$ where $F$ is a final state. The set of all live states is called live complete set denoted $P$ and states which not live are called dead states. In isomorphic automata, there is only one dead state, $d_0$. The set having live states as well as dead state is denoted as $P'$ that is $P = P \cup \{d_0\}$. The ID algorithm partitions the set $T$ into blocks of accepting and non-accepting strings, for this purpose it uses the concept of distinguishing strings $V$. $T$ is a set having all live states as well as their concatenation with input alphabet $\beta$ such as $T = P' \cup \{f(\alpha, \beta) \mid (\alpha, \beta) \in P \times \Sigma\}$ where $\alpha \in P'$ and $\beta \in \Sigma$.

The purpose of distinguishing strings is to identify states, having same behavior For some particular string, $\alpha \in \Sigma^*$ but have different behavior for a suffix $\sigma \in \Sigma$.
To find the blocks of accepting and nonaccepting states, the ID constructs a table. The first row of table shows the number of iterations, through which the set $T$ is partitioned into accepting and nonaccepting blocks. The second row of table shows the set of distinguishing strings $v_1, v_2, \ldots, v_n$ where $v_1, v_2, \ldots, v_n \in V$. First column of table shows the elements of the set $T$ with transition function $E$ where $E_i(\alpha) = \{v_j \mid v_j \in V, 0 \leq j \leq i, \alpha v_j \in L(A) ?\}$ and $L(A)$ is the language of target DFA.

| $i$ | 0 | 1 | 2 |
|---|---|---|---|
| $v_i$ | $\lambda$ | $a$ | $b$ |
| $E(d_0)$ | $\varphi$ | $\varphi$ | $\varphi$ |
| $E(\lambda)$ | | | |
| $E(a)$ | | | |
| $E(b)$ | | | |
| $E(aa)$ | | | |
| $E(ab)$ | | | |
| $E(ba)$ | | | |
| $E(bb)$ | | | |

Table 2.2.1: Structure of Table

Table 2.2.1 shows that two iterations have been completed to reach final partition of blocks and the second row shows that $V = \{ \lambda, a, b \}$. From third row we can see that any transition from dead state is always dead, $E_i(d_0) = \varphi$. In first iteration $E_0$ when $v_0 = \lambda$, $E(d_0) = \varphi$ and $E_0(\alpha) = \lambda$ when $\alpha \in L(A)$. Otherwise $E_0(\alpha) = \varphi$.

When first iteration becomes complete, ID searches for a pair such that $E_i(\alpha) = E_i(\beta)$ but $E_i(f(\alpha, \sigma)) \neq E_i(f(\beta, \sigma))$ whereas $\alpha, \beta \in P$ and $\sigma \in \Sigma$. This expression shows that if the ID algorithm finds a pair from set $P'$ which shows the same behavior i.e. either both $E_i(\alpha)$ and $E_i(\beta)$ lie in the accepting block or both lie in rejecting block and when we concatenate $\alpha$ and $\beta$ with some alphabet $\sigma$ from the input set $\Sigma$ then their behavior changes i.e. one lie in accepting block and other lie in rejecting block. This can give a potential distinguishing string. Then ID chooses some string $\gamma \in E_i(f(\alpha, \sigma)) \oplus E_i(f(\beta, \sigma))$ and a new distinguishing string is defined as $\sigma\gamma$. The ID performs next iteration $i + 1$ to further split the blocks, by reading distinguishing string $\sigma\gamma \in \Sigma$ from all elements of $E_i(\alpha)$. For this, it asks membership queries as $\alpha v_{i+1} \in L(A)$ ? , if the adequate teacher answers as Yes then $E_i(\alpha)$ becomes $E_{i-1}(\alpha) \cup \{v_i\}$, and if it answers as No then $E_i(\alpha)$ is set to $E_{i-1}(\alpha)$.
If the ID algorithm finds no such pair that is $E_i(\alpha) = E_i(\beta)$ but $E_i(f(\alpha, \sigma)) \neq E_i(f(\beta, \sigma))$ then it constructs the hypothesis DFA, $H$ which is isomorphic to the target DFA, $A$.

In hypothesis DFA, $H$, $E_i(\alpha)$ represents the states where $\alpha \in T$. $E(\lambda)$ is initial state and $E_i(\alpha)$ where $\alpha \in T$ and $\lambda \in E_i(\alpha)$ are final states. The transition relation $\delta$ is constructed as $E_i(\alpha) = \varphi$ then self-loop to that state otherwise $\delta(E_i(\alpha), \sigma) = E_i(f(\alpha, \sigma))$.

## Example

An example run of the ID algorithm is given below:

Target *DFA*: $A$= Consecutive even number of a's and all b's. $(b*(aa)*b*)$

Input alphabet: $\Sigma = \{a, b\}$

$P_0 = \{\lambda, a\}$ and $P_0' = \{d_0, \lambda, a\}$ and $T_0'$ becomes as $T_0' = \{d_0, \lambda, a, b, aa, ab\}$

| $i$ | 0 |
|---|---|
| $v_i$ | $\lambda$ |
| $E(d_0)$ | $\varphi$ |
| $E(\lambda)$ | $\{\lambda\}$ |
| $E(a)$ | $\varphi$ |
| $E(b)$ | $\{\lambda\}$ |
| $E(aa)$ | $\{\lambda\}$ |
| $E(ab)$ | $\varphi$ |

Table 2.2.2: Initial Table

Table 2.2.2 shows that distinguishing string set $V = \{\lambda\}$. The ID algorithm asks membership queries for all strings belong to $T$ as $av_{i+1} \in L(A)$. The adequate teacher answers as Yes for $E(\lambda)$, $E(b)$, $E(aa)$ as these strings lead to the accepting states so $E_i(\alpha)$ becomes $\{\lambda\}$ and for all others, those are not leading to accepting states, adequate teacher answers as No so they set to $\varphi$.

Table 2.2.2 shows that $E(d_0) = E(a)$ but $E(d_0.\ \alpha) \neq E(a.\ a)$ therefore we can take a distinguishing string $\sigma\gamma$ as "$a$" in next iteration. The extended table is given in Table 2.2.3

| i | 0 | 1 |
|---|---|---|
| $v_i$ | $\lambda$ | $a$ |
| $E(d_0)$ | $\varphi$ | $\varphi$ |
| $E(\lambda)$ | $\{\lambda\}$ | $\{\lambda\}$ |
| $E(a)$ | $\varphi$ | $\{a\}$ |
| $E(b)$ | $\{\lambda\}$ | $\{\lambda\}$ |
| $E(aa)$ | $\{\lambda\}$ | $\{\lambda\}$ |
| $E(ab)$ | $\varphi$ | $\varphi$ |

Table 2.2.3: For Distinguishing String $a$

Table 2.2.3 shows that distinguishing string set $V = \{\lambda, a\}$. The ID algorithm asks membership queries for all strings belong to $T$ as $av_{i+1} \in L(A)$. The adequate teacher answers as Yes for $E(\lambda)$, $E(a)$, $E(b)$, $E(aa)$ as these strings lead to the accepting states so $E_i(\alpha)$ becomes $E_{i-1}(\alpha) \cup \{v_i\}$ and for all others, those are not leading to accepting states, adequate teacher answers as No so they set to $\varphi$.

Table 2.2.3 shows that $E(\lambda) = E(a)$ but $E(\lambda.\ b)$ not equal to $E(a.\ b)$ therefore we can take a distinguishing string $\sigma\gamma$ as "$b$" in next iteration. The extended table is given in Table 2.2.4.

| i | 0 | 1 | 2 |
|---|---|---|---|
| $v_i$ | $\lambda$ | $a$ | $b$ |
| $E(d_0)$ | $\varphi$ | $\varphi$ | $\varphi$ |
| $E(\lambda)$ | $\{\lambda\}$ | $\{\lambda\}$ | $\{\lambda, b\}$ |
| $E(a)$ | $\varphi$ | $\{a\}$ | $\{a\}$ |
| $E(b)$ | $\{\lambda\}$ | $\{\lambda\}$ | $\{\lambda, b\}$ |
| $E(aa)$ | $\{\lambda\}$ | $\{\lambda\}$ | $\{\lambda, b\}$ |
| $E(ab)$ | $\varphi$ | $\varphi$ | $\varphi$ |

Table 2.2.4: For Distinguishing String $b$

Table 2.2.4 shows that distinguishing string set $V = \{\lambda, a, b\}$. The ID algorithm asks membership queries for all strings belong to $T$ as $av_{i+1} \in L(A)$. The adequate teacher answers as Yes for $E(\lambda), E(a), E(b), E(aa)$ as these strings lead to the accepting states so $E_i(\alpha)$ becomes $E_{i-1}(\alpha) \cup \{v_i\}$ and for all others, those are not leading to accepting states, adequate teacher answers as No so they set to $\varphi$.

Table 2.2.4 shows that there is no pair such that $E_i(\alpha) = E_i(\beta)$ but $E_i(f(\alpha, \sigma)) \mathrel{/=} E_i(f(\beta, \sigma))$ so blocks will not be further partitioned. The hypothesis DFA, $H$ is given below in Figure 2.2.1.

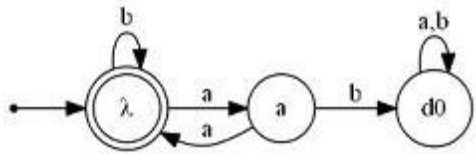

Figure 2.2.1: Hypothesis $H$

As we can see that this automaton is behaviorally equivalent to the target DFA, $A$ i.e. $L(H) = L(A)$ therefore the ID algorithm terminates.

## 2.3 IID Algorithm

The IID is an incremental extension of the ID algorithm [13]. It works similar to the ID algorithm except that it does not require the availability of live complete set at the start of inference procedure. The IID incrementally builds the live state set and its corresponding automata to provided labeled examples, those are taken as input by this algorithm. A labeled example is a pair ($\alpha$, $label(\alpha)$) where $\alpha \in \Sigma^*$ and $label(\alpha)$ shows that whether an equivalence query $\alpha$ is accepted or rejected by the adequate teacher. If $\alpha \in L(A)$ then it is called a positive example and if $\alpha \notin L(A)$ then it is called a negative example. Initial hypothesis DFA, $H_0$ consists of only one state that is the dead state. When the first positive example is seen then $H_0$ is updated and after that for each additional labeled example ($\alpha$, $label(\alpha)$), it is determined that whether it is consistent with our previous hypothesis DFA or we have to update it according to a new labeled example.

Like ID algorithm, IID has a set of live states $P$ that is initially empty. $P'$ is $P' = P \cup \{d_0\}$ where $d_0$ is a dead state. $T'$ is a set having all live states as well as their concatenation with input alphabets $\beta$ such that $T' = P' \cup \{f(\alpha, \beta) \mid (\alpha, \beta) \in P' \times \Sigma\}$ whereas $\alpha \in P'$ and $\beta \in \Sigma$. IID algorithm splits the set $T$ into blocks of accepting and nonaccepting states and for this, it uses the concept of distinguishing strings, denoted by $V$ like ID algorithm. The purpose of distinguishing strings is to identify states, having same behavior for some particular string, $\alpha \in \Sigma^*$ but have different behavior for a suffix $\sigma \in \Sigma$. Like the ID algorithm, to find the blocks of accepting and nonaccepting states, the IID also constructs a table. The first row of table shows the number of iterations, through which the set $T$ is partitioned into accepting and nonaccepting blocks. The second row of table shows the set of distinguishing strings $v_1, v_2, \ldots, v_\eta$ where $v_1, v_2, \ldots, v_n \in V$. First column of table shows the elements of the set $T$ with transition function $E$ where $E_i(\alpha) = \{v_j \mid v_j \in V, 0 \leq j \leq i, \alpha v_j \in L(A)\}$ and $L(A)$ is the language of target DFA.

When the first positive example arrives, the IID algorithm constructs the sets $P, T$ and corresponding table as in the ID algorithm. In the first iteration, the function $E_0$ when $v_0 = \lambda$, $E_0(d_0) = \varphi$ and $E_0(\alpha) = \lambda$ when $\alpha \in L(A)$, otherwise $E_0(\alpha) = \varphi$. After that IID searches for a pair such that $E_i(\alpha) = E_i(\beta)$ but $E_i(f(\alpha, \sigma)) \neq E_i(f(\beta, \sigma))$ whereas $\alpha, \beta \in P$ and $\sigma \in \Sigma$. This expression shows that if IID algorithm finds a pair from set $P'$ which shows the same behavior i.e. either both $E_i(\alpha)$ and $E_i(\beta)$ lie in the accepting block or both lie in rejecting block and when we concatenate $\alpha$ and $\beta$ with some alphabet $\sigma$ from the input set $\Sigma$ then their behavior may change i.e. one lie in accepting block and other lie in rejecting block. This can give a potential distinguishing string. Then IID non-deterministically chooses some string $\gamma \in E_i(f(\alpha, \sigma)) \oplus E_i(f(\beta, \sigma))$ and the new distinguishing string is defined as $\sigma\gamma$. IID performs next iteration $i + 1$ to further split the blocks, by reading distinguishing string $\sigma\gamma \in \Sigma$ from all elements of $E_i(\alpha)$. For this it asks membership queries as $\alpha v_{i+1} \in L(A)$, if adequate teacher answers as Yes then $E_i(\alpha)$ becomes $E_{i-1}(\alpha) \cup \{v_i\}$, and if it replies as No then $E_i(\alpha)$ is set to $E_{i-1}(\alpha)$. If IID finds no such pair that is $E_i(\alpha) = E_i(\beta)$ but $E_i(f(\alpha, \sigma)) \neq E_i(f(\beta, \sigma))$ then it construct the hypothesis DFA, $H_m$. If $H_m$ becomes behaviorally equal to the target DFA, $A$ then IID terminates otherwise, it waits for another positive example and above process repeats until hypothesis DFA, $H$ becomes equivalent to the target DFA, $A$.

## Example

An example run of the IID algorithm is given below:

Target $DFA$: $A=$ Consecutive even number of a's and all b's. ($b^*(aa)^*b^*$)

Input alphabet: $\Sigma = \{a, b\}$

Initial null automata $H_0$ is given below in Figure 2.3.1.

$P_0 = \{\ \}$ and $P_0' = \{d_0\}$ and $T_0'$ becomes as $T_0' = \{d_0, \lambda, a, b\}$

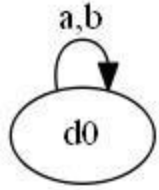

Figure 2.3.1: Null Hypothesis

Suppose the rst labeled example is $(a, -)$, as it is negative example and is consistent with $H_0$ so $H_0$ does not change.

Suppose next labeled example is $(b, +)$ as it is a first positive example so set $P_1 = \{\lambda, b\}$, $P_1 = \{d_0, \lambda, b\}$ and $T_1 = \{d_0, \lambda, a, b, ba, bb\}$

For $k = 0$

| i | 0 |
|---|---|
| vi | λ |
| E(do) | φ |
| E(λ) | {λ} |
| E(a) | φ |
| E(b) | {λ} |
| E(ba) | φ |
| E(bb) | {λ} |

Table 2.3.1: For Distinguishing String λ

Table 2.3.1 shows that distinguishing string set $V = \{\lambda\}$. The IID algorithm asks membership queries for all strings belong to $T$ as $av_{i+1} \in L(A)$. The adequate teacher answers as Yes for $E(\lambda)$, $E(b)$, $E(bb)$ as these strings lead to the accepting states so $E_i(\alpha)$ becomes $\{\lambda\}$ and for all others, those are not leading to accepting states, adequate teacher replies as No so they set to $\varphi$.

Table 2.3.1 shows that there is no pair such that $E_i(\alpha) = E_i(\beta)$ but $E_i(f(\alpha, \sigma)) \neq E_i(f(\beta, \sigma))$ so blocks will not be further partitioned. The hypothesis DFA, $H_1$ for this iteration is given below in Figure 2.3.2.

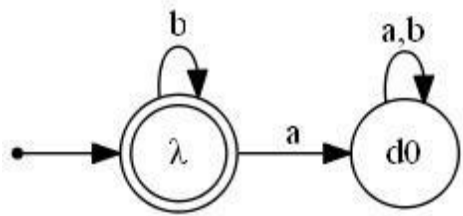

Figure 2.3.2: Hypothesis $H_1$

As automaton $H_1$ represented in Figure 2.3.2 is not equal to the target DFA, $A$ therefore the IID algorithm learns the target DFA with more labeled example.

Suppose next input labeled example is $(aa, +)$

Now $P_2 = \{\lambda, a, b, aa\}$

$P_2' = \{d_0, \lambda, a, b, aa\}$

$T_1' = \{d_0, \lambda, a, b, aa, bb, ab, ba, aaa, aab\}$

$k = 1$

| i | 0 | 1 |
|---|---|---|
| $v_i$ | $\lambda$ | $a$ |
| $E(d_0)$ | $\varphi$ | $\varphi$ |
| $E(\lambda)$ | $\{\lambda\}$ | $\{\lambda\}$ |
| $E(a)$ | $\varphi$ | $\{a\}$ |
| $E(b)$ | $\{\lambda\}$ | $\{\lambda\}$ |
| $E(aa)$ | $\{\lambda\}$ | $\{\lambda\}$ |
| $E(ab)$ | $\varphi$ | $\varphi$ |
| $E(ba)$ | $\varphi$ | $\{a\}$ |
| $E(bb)$ | $\{\lambda\}$ | $\{\lambda\}$ |
| $E(aaa)$ | $\varphi$ | $\{a\}$ |
| $E(aab)$ | $\{\lambda\}$ | $\{\lambda\}$ |

Table 2.3.2: For Distinguishing String $a$

In Table 2.3.2, the column $\lambda$ shows that as $E(d_0) = \varphi = E(a)$ but $E(d_0.a) \neq E(a.a)$ so the IID algorithm partitions the accepting and nonaccepting blocks by using distinguishing string $\sigma\gamma = a$ shown in column $a$ of Table 2.3.2.

Table 2.3.2 shows that distinguishing string set $V = \{\lambda, a\}$. The IID asks membership queries for all strings belong to $T$ as $\alpha v_{i+1} \in L(A)$. The adequate teacher answers as Yes for $E(a)$, $E(ba)$ and $E(aaa)$ as these strings lead to the accepting states so $E_i(\alpha)$ becomes $E_{i-1}(\alpha) \cup \{v_i\}$ and for all others, those are not leading to accepting states, adequate teacher replies as No so they set to $E_{i-1}(\alpha)$.

Table 2.3.2 shows that there is no pair such that $E_i(\alpha) = E_i(\beta)$ but $E_i(f(\alpha, \sigma)) \neq E_i(f(\beta, \sigma))$ so blocks will not be further partitioned. Hypothesis DFA, $H_2$ is given below in Figure 2.3.3.

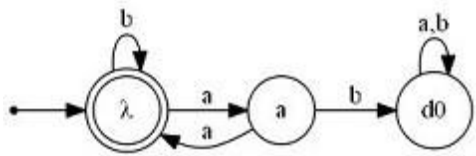

Figure 2.3.3: Hypothesis $H_2$

As above hypothesis DFA, $H_2$ is behaviorally equivalent to the target DFA, $A$ i.e. $L(H_2) = L(A)$ therefore the IID algorithm terminates.

## 2.4 IDS Algorithm

The IDS algorithm is also an incremental extension of ID algorithm [11]. Like IID algorithm, it also does not require the availability of live complete set at the start of inference procedure. IDS incrementally build the live state set and its corresponding automata regarding to provide input labeled examples. A labeled example consists of two parts such as ($\alpha$, *label*($\alpha$)) whereas $\alpha \in \Sigma^*$ and *label*($\alpha$) shows that whether an equivalence query $\alpha$ is accepted or rejected by the adequate teacher. A *label*($\alpha$) is valued as accepted if $\alpha \in L(A)$ and *label*($\alpha$) is valued as rejected if $\alpha \notin L(A)$. If $\alpha \in L(A)$ then it is called positive example and if $\alpha \notin L(A)$ then it is called negative example. Initial hypothesis *DFA*, $H_0$ consists of only one state (initial state) and all its input alphabet transitions $\beta \in \Sigma$. When a first labeled example (either positive or negative) arrives, then $H_0$ is updated and after that for each additional labeled example ($\alpha$, *label*($\alpha$)), it is determined that whether it is consistent with our previous hypothesis *DFA* or we have to update it according to new labeled example.

Like IID algorithm [13], the IDS algorithm has also a set $P$ that is initially as $P = \{\lambda\}$ and $P' = P \cup \{d_0\}$, where $d_0$ is a dead state. $T'$ is a set having all states as well as their concatenation with input alphabet $\beta$ such that $T' = P' \cup \{f(\alpha, \beta) \mid (\alpha, \beta) \in P' \times \Sigma\}$ where $\alpha \in P'$ and $\beta \in \Sigma$ (for prefix closed). The IDS algorithm partitions the set $T'$ into the blocks of accepting and nonaccepting states and for this, it uses the concept of distinguishing strings $V$ like the ID and IID algorithms. The purpose of distinguishing strings is to identify a string, having same behavior for some particular string, $\alpha \in \Sigma^*$ but have different behavior for a suffix $\sigma \in \Sigma$.

Like ID and IID algorithms, to find the blocks of accepting and nonaccepting states, the IDS algorithm also constructs a table. The first row of table shows the number of iterations, through which set $T'$ is partitioned into accepting and nonaccepting blocks. The second row of table shows the set of distinguishing strings $v_1, v_2, \ldots, v_n$ where $v_1, v_2, \ldots, v_n \in V$. First column of table shows the elements of the set $T'$ with transition function $E$ where $E_i(\alpha) = \{v_j | v_j \in V, 0 \leq j \leq i, \alpha v_j \in L(A)?\}$ and $L(A)$ is the language of target DFA, $A$.

When IDS receives a first labeled example, it constructs the set $P$, $P'$, $T'$ and corresponding table like in ID algorithm. In first iteration $E_0$ when $v_0 = \lambda$, $E(d_0) = \varphi$ and $E_0(\alpha) = \lambda$ when $\alpha \in L(A)$. Otherwise $E_0(\alpha) = \varphi$. After that IDS searches for a pair such that $E_i(\alpha) = E_i(\beta)$ but $E_i(f(\alpha, \sigma)) \neq E_i(f(\beta, \sigma))$ whereas $\alpha, \beta \in P'$ and $\sigma \in \Sigma$. This expression shows that if IDS finds a pair from set $P$ which shows the same behavior i.e. either both $E_i(\alpha)$ and $E_i(\beta)$ lie in accepting block or both lie in rejecting block and when we concatenate $E_i(\alpha)$ and $E_i(\beta)$ with some alphabet $\sigma$ from the input set $\Sigma$ then their behavior changes i.e. one lie in accepting block and other lie in rejecting block. Then the IDS algorithm chooses some string $\gamma \in E_i(f(\alpha, \sigma)) \oplus E_i(f(\beta, \sigma))$ and a new distinguishing string is defined as $\sigma\gamma$. IDS perform next iteration $i + 1$ to further split the blocks, by reading distinguishing string $\sigma\gamma$
$\in \Sigma$ from all elements of $E_i(\alpha)$. For this, it asks membership queries as $\alpha v_i +1$
$\in L(A)$? if the adequate teacher answers as Yes then $E_i(\alpha)$ becomes $E_{i-1}(\alpha)$
$\cup \{v_i\}$, if the adequate teacher answers as No then $E_i(\alpha)$ is set to $E_{i-1}(\alpha)$. If the IDS finds no such pair that is $E_i(\alpha) = E_i(\beta)$ but $E_i(f(\alpha, \sigma)) \neq E_i(f(\beta, \sigma))$ then it constructs the hypothesis DFA $H_m$. If $H_m$ is equivalent to the target DFA, $A$ then the IDS algorithm stops its execution otherwise it waits for another labeled example. Above process repeats until hypothesis DFA, $H$ becomes equivalent to the target DFA, $A$.

This algorithm has two versions. One is prefix closed like L*, ID, IID and other one is prefix free. The main difference between these two is; in prefix free version, set $P$ contains the strings gained from the labeled examples without their prefixes whereas, in prefix closed version, set $P$ contains the strings gained from labeled examples as well as their prefixes.

## Example

### Prefix Closed

An example run of prefix closed version of IDS algorithm is given below:
Input alphabet: $\Sigma = \{a, b\}$
Target DFA: $A$= Consecutive even number of a's and all b's. $(b*(aa)*b*)$

Initially $P_0 = \{\lambda\}$, $P'_0 = \{d_0, \lambda\}$
$T'_0 = \{d_0, \lambda, a, b\}$
Initial null automata $H_0$ is given below in Figure 2.4.1.

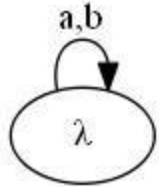

Figure 2.4.1: Null Automaton $H_0$

Suppose the first labeled example is $(a, -)$, as this example shows that transition $a$ from initial state $\lambda$ leads to the dead state so it is not consistent with $H_0$. Here $P_1 = \{\lambda, a\}$, $P'_1 = \{d_0, \lambda, a\}$ and $T'_1 = \{d_0, \lambda, a, b, aa, ab\}$ so Table 2.4.1 is given below:

| i | 0 |
|---|---|
| Vi | λ |
| E(d₀) | φ |

| $E(\lambda)$ | $\{\lambda\}$ |
|---|---|
| $E(a)$ | $\varphi$ |
| $E(b)$ | $\{\lambda\}$ |
| $E(aa)$ | $\varphi$ |
| $E(ab)$ | $\varphi$ |

Table 2.4.1: For Labeled Example ($a$, -)

Table 2.4.1 shows that distinguishing string set $V=\{\lambda\}$ and IDS asks membership queries for all strings belong to $T$ as $av_{i+1}\in L(A)$. The adequate teacher answers Yes, for $E(\lambda)$ and $E(b)$ as these strings lead to the accepting states so $E_0(\alpha)$ becomes $\{\lambda\}$ and for all others, those are not leading to accepting states, adequate teacher replies No so they are set to $\varphi$.

Table 2.4.1 shows that there is no pair such that $E_i(\alpha) = E_i(\beta)$ but $E_i(f(\alpha, \sigma))$ /= $E_i(f(\beta, \sigma))$ so blocks will not be further partitioned. Hypothesis $DFA$, $H_1$ is given below in Figure 2.4.2.

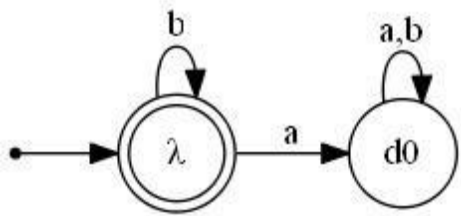

Figure 2.4.2: Hypothesis $H_1$

Suppose next labeled example is ($ab$,—) so $P_2 = \{\lambda, a, ab\}$ and $P'_2 = \{d_0, \lambda, a, ab\}$ and $T'_2$ becomes as $T'_2 = \{d_0, \lambda, a, b, aa, ab, aba, abb\}$

The corresponding table for this iteration is given below:

| $i$ | 0 |
|---|---|
| $v_i$ | $\Lambda$ |
| $E(d_0)$ | $\Phi$ |
| $E(\lambda)$ | $\{\lambda\}$ |
| $E(a)$ | $\Phi$ |
| $E(b)$ | $\{\lambda\}$ |
| $E(aa)$ | $\Phi$ |
| $E(ab)$ | $\Phi$ |
| $E(aba)$ | $\Phi$ |
| $E(abb)$ | $\Phi$ |

Table 2.4.2: For Labeled Example ($ab$, -)

Table 2.4.2 shows that distinguishing string set $V=\{\lambda\}$ and IDS asks membership queries for all strings belong to $T$ as $av_{i+1}\in L(A)$. The adequate teacher answers Yes for $E(\lambda)$ and $E(b)$ as these strings lead to the accepting states so $E_0(\alpha)$ becomes $\{\lambda\}$ and for all others, those are not leading to accepting states, adequate teacher replies No so they are set to $\varphi$.

Table 2.4.2 shows that there is no pair such that $E_i(\alpha) = E_i(\beta)$ but $E_i(f(\alpha, \sigma)) \neq E_i(f(\beta, \sigma))$ so blocks will not be further refined. Hypothesis DFA, $H_2$ is given below in Figure 2.4.3

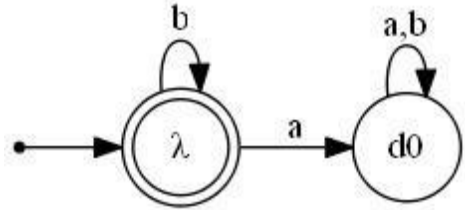

Figure 2.4.3: Hypothesis $H_2$

Suppose next labeled example is $(b,+)$ so $P_3 = \{\lambda, a, b, ab\}$ and $P_3' = \{d_0, \lambda, a, b, ab\}$ and $T'$ becomes as $T_3' = \{d_0, \lambda, a, b, aa, ab, ba, bb, aba, abb\}$

The corresponding table for the current iteration is given below:

| i | 0 |
|---|---|
| vi | $\lambda$ |
| E(do) | $\varphi$ |
| E($\lambda$) | $\{\lambda\}$ |
| E(a) | $\varphi$ |
| E(b) | $\{\lambda\}$ |
| E(aa) | $\varphi$ |
| E(ab) | $\varphi$ |
| E(ba) | $\varphi$ |
| E(bb) | $\{\lambda\}$ |
| E(aba) | $\varphi$ |
| E(abb) | $\varphi$ |

Table 2.4.3: For Labeled Example $(b, +)$

Table 2.4.3 shows that distinguishing string set $V=\{\lambda\}$ and IDS asks membership queries for all strings belong to $T$ as $av_{i+1} \in L(A)$. The adequate teacher answers Yes for $E(\lambda)$, $E(b)$ and $E(bb)$ as these strings lead to the accepting states so $E_0(\alpha)$ becomes $\{\lambda\}$ and for all others, those are not leading to accepting states, adequate teacher replies No so they are set to $\varphi$.

Table 2.4.3 shows that there is no pair such that $E_i(\alpha) = E_i(\beta)$ but $E_i(f(\alpha, \sigma)) /= E_i(f(\beta, \sigma))$ so blocks will not be further partitioned. Hypothesis DFA, $H_3$ is given below in Figure 2.4.4.

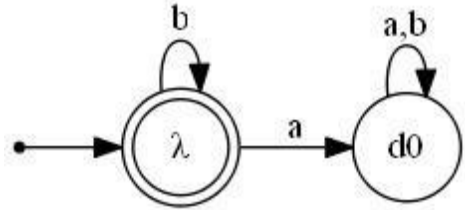

Figure 2.4.4: Hypothesis $H_3$

Suppose next labeled example is $(aa, +)$
Now $P_4 = \{\lambda, a, b, aa, ab\}$
$P_4' = \{d_0, \lambda, a, b, aa, ab\}$
$T_4' = \{d_0, \lambda, a, b, aa, bb, ab, ba, aaa, aab, aba, abb\}$
therefore Table 2.4.4 is given below:

| i | 0 | 1 |
|---|---|---|
| vi | $\lambda$ | a |
| E(do) | $\varphi$ | $\varphi$ |
| E($\lambda$) | $\{\lambda\}$ | $\{\lambda\}$ |
| E(a) | $\varphi$ | $\{a\}$ |
| E(b) | $\{\lambda\}$ | $\{\lambda\}$ |
| E(aa) | $\{\lambda\}$ | $\{\lambda\}$ |
| E(ab) | $\varphi$ | $\varphi$ |
| E(ba) | $\varphi$ | $\{a\}$ |
| E(bb) | $\{\lambda\}$ | $\{\lambda\}$ |
| E(aaa) | $\varphi$ | $\{a\}$ |
| E(aab) | $\{\lambda\}$ | $\{\lambda\}$ |
| E(aba) | $\varphi$ | $\varphi$ |
| E(abb) | $\varphi$ | $\varphi$ |

Table 2.4.4: For Labeled Example $(aa, +)$

In Table 2.4.4, the column $\lambda$ shows that as $E(d_0) = \varphi = E(a)$ but $E(d_0.a) \neq E(a.a)$ hence the IDS algorithm partitions the accepting and nonaccepting blocks by using distinguishing string $\sigma\gamma = a$ shown in column $a$ of Table 2.4.4.

Table 2.4.4 shows that distinguishing string set $V = \{\lambda, a\}$. The IDS asks membership queries for all strings belong to $T$ as $av_{i+1} \in L(A)$. The adequate teacher answers Yes for $E(a)$, $E(ba)$ and $E(aaa)$ as these strings lead to the accepting states so $E_i(\alpha)$ becomes $E_{i-1}(\alpha) \cup \{v_i\}$ and for all others, those are not leading to accepting states, adequate teacher replies No so they are set to $E_{i-1}(\alpha)$.

Table 2.4.4 shows that there is no pair such that $E_i(\alpha) = E_i(\beta)$ but $E_i(f(\alpha, \sigma)) \neq E_i(f(\beta, \sigma))$ so blocks will not be further partitioned. Hypothesis DFA, $H_4$ is given below in Figure 2.4.5.

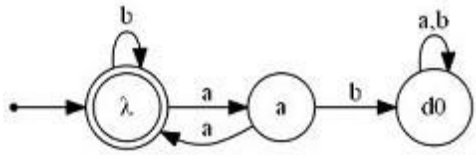

Figure 2.4.5: Hypothesis $H_4$

As the above hypothesis *DFA*, $H_4$ is behaviorally equivalent to the target DFA, $A$ i.e. $L(H_4) = L(A)$ therefore the IDS algorithm terminates.

Prefix Free

An example run of prefix free version of IDS algorithm is given below:
Target *DFA*: *A*= Consecutive even number of a's and all b's. $(b*(aa)*b*)$
Input alphabet: $\Sigma = \{a, b\}$

Initially $P_0 = \{\lambda\}$, $P'_0 = \{d_0, \lambda\}$
$T'_0 = \{d_0, \lambda, a, b\}$
Initial null automata $H_0$ is given below in Figure 2.4.6.

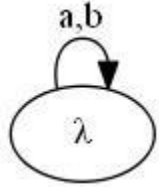

Figure 2.4.6: Null Hypothesis $H_0$

Suppose the first labeled example is $(ab, —)$ therefore
$P_1 = \{\lambda, ab\}$, $P'_1 = \{d_0, \lambda, ab\}$ and $T'_1 = \{d_0, \lambda, a, b, ab, aba, abb\}$ so

corresponding Table 2.4.5 is given below:

| i | 0 |
|---|---|
| vi | $\lambda$ |
| $E(d_0)$ | $\varphi$ |
| $E(\lambda)$ | $\{\lambda\}$ |
| $E(a)$ | $\varphi$ |
| $E(b)$ | $\{\lambda\}$ |
| $E(ab)$ | $\varphi$ |
| $E(aba)$ | $\varphi$ |
| $E(abb)$ | $\varphi$ |

Table 2.4.5: For Labeled Example $(ab, -)$

Table 2.4.5 shows that distinguishing string set $V = \{\lambda\}$ and IDS asks membership queries for all strings belong to $T$ as $av_{i+1} \in L(A)$. The adequate teacher answers Yes for $E(\lambda)$ and $E(b)$ as these strings lead to the accepting states so $E_0(\alpha)$ becomes $\{\lambda\}$ and for all others, those are not leading to accepting states, adequate teacher replies No so they are set to $\varphi$.

Table 2.4.5 shows that there is no pair such that $E_i(\alpha) = E_i(\beta)$ but $E_i(f(\alpha, \sigma)) \ne E_i(f(\beta, \sigma))$ so blocks will not be further partitioned. Hypothesis *DFA*, $H_1$ is given below in Figure 2.4.7.

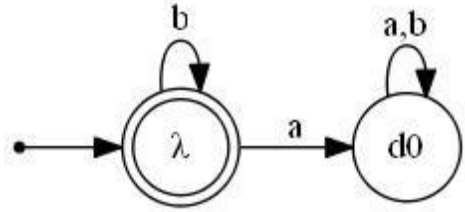

Figure 2.4.7: Hypothesis $H_1$

Suppose next labeled example is $(b, +)$ so
$P_2 = \{\lambda, b, ab\}$ and $P_2' = \{d_0, \lambda, b, ab\}$ and $T_2'$ becomes as $T_2' = \{d_0, \lambda, a, b, ab, ba, bb, aba, abb\}$
The corresponding table for the current iteration is given below:

| i | 0 |
|---|---|
| $v_i$ | $\lambda$ |
| $E(d_0)$ | $\varphi$ |
| $E(\lambda)$ | $\{\lambda\}$ |
| $E(a)$ | $\varphi$ |
| $E(b)$ | $\{\lambda\}$ |
| $E(ab)$ | $\varphi$ |
| $E(ba)$ | $\varphi$ |
| $E(bb)$ | $\{\lambda\}$ |
| $E(aba)$ | $\varphi$ |
| $E(abb)$ | $\varphi$ |

Table 2.4.6: For Labeled Example $(b, +)$

Table 2.4.6 shows that distinguishing string set $V = \{\lambda\}$ and IDS asks membership queries for all strings belong to $T$ as $av_{i+1} \in L(A)$. The adequate teacher answers Yes for $E(\lambda)$, $E(b)$ and $E(bb)$ as these strings lead to the accepting states so $E_0(\alpha)$ becomes $\{\lambda\}$ and for all others, those are not leading to accepting states, adequate teacher replies No so they are set to $\varphi$.

Table 2.4.6 shows that there is no pair such that $E_i(\alpha) = E_i(\beta)$ but $E_i(f(\alpha, \sigma)) /= E_i(f(\beta, \sigma))$ so blocks will not be further partitioned. Hypothesis DFA, $H_2$ is given below in Figure 2.4.8.

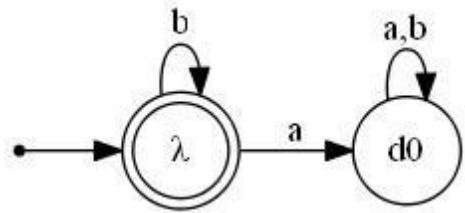

Figure 2.4.8: Hypothesis $H_2$

Suppose next labeled example is (*aa, +*)
Now $P_3$ = {λ, *b, aa, ab*}
$P_3'$ = {$d_0$, λ, *b, aa, ab*}
$T_3'$ = {$d_0$, λ, *a, b, aa, bb, ab, ba, aaa, aab, aba, abb*}
therefore Table 2.4.7 is given below:

| i | 0 | 1 |
|---|---|---|
| vi | λ | a |
| E(do) | φ | φ |
| E(λ) | {λ} | {λ} |
| E(a) | φ | {a} |
| E(b) | {λ} | {λ} |
| E(aa) | {λ} | {λ} |
| E(ab) | φ | φ |
| E(ba) | φ | {a} |
| E(bb) | {λ} | {λ} |
| E(aaa) | φ | {a} |
| E(aab) | {λ} | {λ} |
| E(aba) | φ | φ |
| E(abb) | φ | φ |

Table 2.4.7: For Labeled Example (*aa, +* )

In Table 2.4.7, the column λ shows that as $E(a) = φ = E(ab)$ but $E(a. a) ≠ E(ab. a)$ so the IDS algorithm partitions the accepting and nonaccepting blocks by using distinguishing string $σγ = a$ shown in column *a* of table 2.4.7.

Table 2.4.7 shows that distinguishing string set $V = \{λ, a\}$. The IDS asks membership queries for all strings belong to $T$ as $av_{i+1} \in L(A)$. The adequate teacher answers Yes for $E(a)$, $E(ba)$ and $E(aaa)$ as these strings lead to the accepting states so $E_i(α)$ becomes $E_{i-1}(α) ∪ \{v_i\}$ and for all others, those are not leading to accepting states, adequate teacher replies No so they are set to $E_{i-1}(α)$.

Table 2.4.7 shows that there is no pair such that $E_i(α) = E_i(β)$ but $E_i(f(α, σ)) /= E_i(f(β, σ))$ so blocks will not be further partitioned. Hypothesis DFA, $H_3$ is given below in Figure 2.4.9.

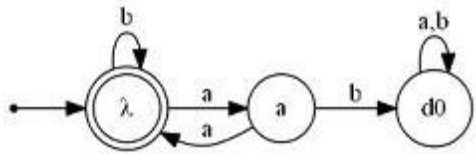

Figure 2.4.9: Hypothesis $H_3$

As above hypothesis DFA, $H_3$ is behaviorally equivalent to the target DFA, *A* i.e. $L(H_3) = L(A)$ therefore the IDS algorithm terminates.

## 2.5 IKL Algorithm

IKL is an incremental extension of the ID algorithm like the IDS algorithm but the major difference is, the IID algorithm is one bit whereas the IKL is a multibit extension [12]. The IKL learns deterministic Kripke structures multi-bit moore machine having k-bit outputs. The IKL algorithm uses two basic ideas. One is bit-slicing i.e. converting k-bits Kripke structure into k 1-bit Kripke structures having 1-bit output which is given below in Figure 2.51. Second concept used is of *partition refinement* which is similar to the consistency maintanence of the ID algorithm but difference is, the IKL algorithm uses the concept of lazy partition refinement.

Like IDS algorithm, the IKL algorithm has also a set $P$ that is initially as $P_0 = \{\varepsilon\}$ and $P'_0 = P_0 \cup \{d_0\}$, where $d_0$ is a dead state. $T'$ is a set having all states as well as their concatenation with input alphabet $\beta$ such that $T_k = T_{k-1} \cup P' \cup \{(\alpha, \beta) \mid (\alpha \in P_k - P_{k-1}, \beta \in \Sigma\}$ for pre x closure. The IKL algorithm partitions the set $T'$ into the blocks of accepting and nonaccepting states and for this, it uses the concept of distinguishing strings $V$ like the ID and IID algorithms. The purpose of distinguishing strings is to identify a state, having same behavior for some particular string, $\alpha \in \Sigma^*$ but have different behavior for a suffix $\sigma \in \Sigma$. Like ID and IDS algorithms, to find the blocks of accepting and nonaccepting states, the IDS algorithm also constructs a table. The  rst row of table shows the number of iterations, through which set $T'$ is partitioned into accepting and nonaccepting blocks. The second row of table shows the set of distinguishing strings $v_1, v_2, \ldots, v_n$ where $v_1, v_2, \ldots, v_n \in V$. First column of table shows the elements of the set $T'$ with transition function $E$ where $E^c(\alpha) = \{v_j \mid v_j \in V, 0 \leq j \leq i, \alpha v_j \in L(A)?\}$
and $L(A)$ is the language of target *DFA*, $A$.

In the IKL algorithm, the target automaton $A$ is initially converted into k 1-bit automata i.e. $B_1, B_2... B_n$ by bit slicing. After that all 1-bit automata are incrementally learned and then the IKL algorithm finds the product of all 1-bit automata $B_1, B_2... B_n$ to convert all 1-bit Kripke structures into k-bit target automata $A$.

The IKL constructs the set $P$, $P'$, $T'$ and corresponding tables for all 1 bit automata $B_1, B_2... B_n$ like in ID algorithm. In first iteration of all tables $E_0$ when $v_0 = \varepsilon$, $E(d_0) = \varphi$ and $E_0(\alpha) = \varepsilon$ when $\alpha \in L(A)$. Otherwise $E_0(\alpha) = \varphi$. After that IKL searches for a pair such that $E_{ic}^c(\alpha) = E_{ic}^c(\beta)$ but $E_{ic}^c(f(\alpha, \sigma)) \neq E_{ic}^c(f(\beta, \sigma))$ whereas $\alpha, \beta \in P'$ and $\sigma \in \Sigma$. This expression shows that if IKL finds a pair from set $P'$ which shows the same behavior i.e. either both $E_{ic}^c(\alpha)$ and $E_{ic}^c(\beta)$ accepted both lie in rejected block and when we concatenate $E_{ic}^c(\alpha)$ and $E_{ic}^c(\beta)$ with some alphabet $\sigma$ from the input set $\Sigma$ then their behavior changes i.e. one accepted block and other is rejected. Then the IKL algorithm chooses some string $\gamma \in E^c(f(\alpha, \sigma)) \oplus E^c(f(\beta, \sigma))$ and a new distinguishing string is defined as $\sigma\gamma$. The IKL algorithm performs next iteration $i + 1$ to further partition the blocks, by reading distinguishing string $\sigma\gamma \in \Sigma$ from all elements of $E_i^c(\alpha)$. For this, it asks membership queries as $\alpha v_i +1 \in L(A)$?

If the adequate teacher answers as Yes then $E^c_{i_c}(\alpha)$ becomes $E^c_{i_{c-1}}(\alpha) \cup \{v_i\}$, if the adequate teacher answers as No then $E^c_{i_c}(\alpha)$ is set to $E^c_{i_{c-1}}(\alpha)$. The IKL repeats the above process until all tables corresponding to $B_1, B_2 ... B_n$ become consistent. After that it constructs the product automata $H_m$. If $H_m$ is behaviorally equivalent to the target automata $A$ and input string set $S$ is empty then the IKL algorithm stops its execution.

### Example

An example run of the IKL algorithm is given below:
Kripke structure = 3 bits
$\Sigma = \{a, b\}$
File S contain = $a, ba$
Target automata $A$ = Odd number of a's

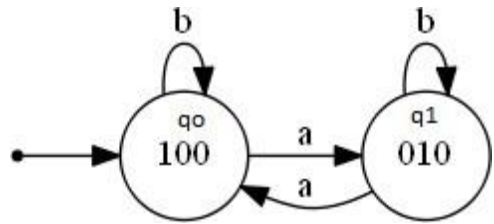

Figure 2.5.1: Target Automata

In the first step, the IKL algorithm finds the bit slicing of the target automata $A$ in form of $B_1, B_2, B_3$ which are given below in Figure 2.5.2.

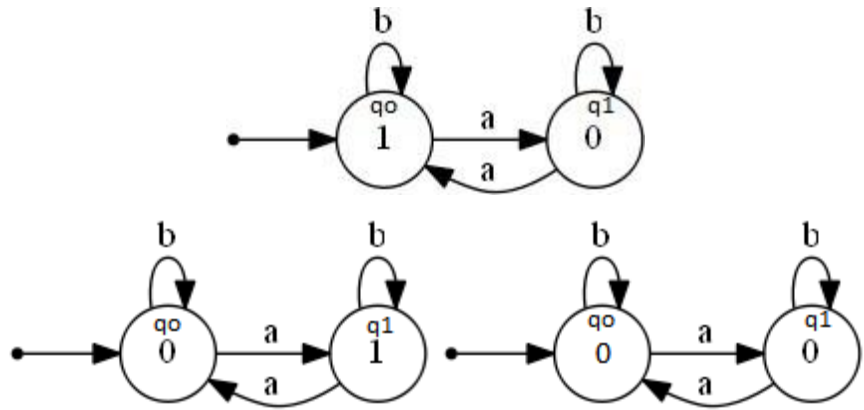

Figure 2.5.2: $B_1, B_2, B_3$

c = 1 → 3
$i_1$ = 0

$i_2 = 0$
$i_3 = 0$
$V_1 = \varepsilon$
$V_2 = \varepsilon$
$V_3 = \varepsilon$
$k = 0, t = 0$
$P_0 = \{\varepsilon\}$
$P'_0 = P_0 \cup \{d_0\} = \{\varepsilon, d_0\}$
$T_0 = P_0 \cup \Sigma = \{\varepsilon, a, b\}$
$E^1_0(d_0) = \varphi$
$E^2_0(d_0) = \varphi$
$E^3_0(d_0) = \varphi$
Now suppose the IKL algorithm reads the input string $a$ then:
$k = 1, t = 1$
$P_1 = P_0 \cup Pref(\alpha) = \{\varepsilon, a\}$
$T_1 = T_0 \cup Pref(\alpha) \cup \{\alpha.b\} = \{\varepsilon, a, b\} \cup \{\varepsilon\} \cup \{a, aa, ab\} = \{\varepsilon, a, b, aa, ab\}$
$T_1 = \{d_0, \varepsilon, a, b, aa, ab\}$

| $i_1$ | 0 | $i_2$ | 0 | $i_3$ | 0 |
|---|---|---|---|---|---|
| $v_1$ | $\varepsilon$ | $v_2$ | $\varepsilon$ | $v_3$ | $\varepsilon$ |
| $E^1(d_0)$ | $\varphi$ | $E^2(d_0)$ | $\varphi$ | $E^3(d_0)$ | $\varphi$ |
| $E^1(\varepsilon)$ | $\varphi$ | $E^2(\varepsilon)$ | $\varphi$ | $E^3(\varepsilon)$ | $\varphi$ |
| $E^1(a)$ | $\varphi$ | $E^2(a)$ | $\{\varepsilon\}$ | $E^3(a)$ | $\varphi$ |
| $E^1(b)$ | $\{\varepsilon\}$ | $E^2(b)$ | $\varphi$ | $E^3(b)$ | $\varphi$ |
| $E^1(aa)$ | $\{\varepsilon\}$ | $E^2(aa)$ | $\varphi$ | $E^3(aa)$ | $\varphi$ |
| $E^1(ab)$ | $\varphi$ | $E^2(ab)$ | $\{\varepsilon\}$ | $E^3(ab)$ | $\varphi$ |

Table 2.5.1: For $B1$    Table 2.5.2: For $B2$    Table 2.5.3: For $B3$

Table 2.5.1 shows that $E^1_0(\varepsilon) = E^1_0(a)$ but $E^1_0(\varepsilon. a) \neq E^1_0(a. a)$ so $a$ is a distinguishing string for Table 2.5.1. Table 2.5.2 shows that $E^1_0(\varepsilon) = E^1_0(d_0)$ but $E^1_0(\varepsilon. a) \neq E^1_0(d_0. a)$ so $a$ is a distinguishing string for Table 2.5.2. Whereas in Table 2.5.3, the corresponding values for all strings belonging to $T'$ are $\varphi$ so it shows that it is consistent and having only one state that is $\varphi$ denoted as $q_0$ for the Table 2.5.3.

Now IKL maintain the consistency of Table 2.5.1 and Table 2.5.2 by updating these table with distinguishing string $a$.

Updated Table 2.5.1 denoted as Table 2.5.1(a) and updated Table 2.5.2 denoted as Table 2.5.2(a) are given below:

| $i_1$ | 0 | 1 |
|---|---|---|
| $v_1$ | $\varepsilon$ | $a$ |
| $E^1(d_0)$ | $\varphi$ | $\varphi$ |
| $E^1(\varepsilon)$ | $\varphi$ | $\varphi$ |
| $E^1(a)$ | $\varphi$ | $\{a\}$ |
| $E^1(b)$ | $\{\varepsilon\}$ | $\{\varepsilon\}$ |
| $E^1(aa)$ | $\{\varepsilon\}$ | $\{\varepsilon\}$ |
| $E^1(ab)$ | $\varphi$ | $\{a\}$ |

| $i_2$ | 0 | 1 |
|---|---|---|
| $v_2$ | $\varepsilon$ | $a$ |
| $E^2(d_0)$ | $\varphi$ | $\varphi$ |
| $E^2(\varepsilon)$ | $\varphi$ | $\{a\}$ |
| $E^2(a)$ | $\{\varepsilon\}$ | $\{\varepsilon\}$ |
| $E^2(b)$ | $\varphi$ | $\{a\}$ |
| $E^2(aa)$ | $\varphi$ | $\{a\}$ |
| $E^2(ab)$ | $\{\varepsilon\}$ | $\{\varepsilon\}$ |

Table 2.5.1(a): For $B_1$     Table 2.5.2(a): For $B_2$

Table 2.5.1(a) shows that $E^1_0(\varepsilon) = E^1_0(d_0)$ but $E^1_0(\varepsilon.\ b) \mathrel{/}= E^1_0(d_0.\ b)$ so $b$ is a distinguishing string for Table 2.5.1(a). Table 2.5.2(a) is now consistent as it finds no such pair that is $E_i(\alpha) = E_i(\beta)$ but $E_i(f(\alpha, \sigma)) \mathrel{/}= E_i(f(\beta, \sigma))$. Updated Table 2.5.1(a) denoted as Table 2.5.1(a') is given below:

| $i_1$ | 0 | 1 | 2 |
|---|---|---|---|
| $v_1$ | $\varepsilon$ | $A$ | $b$ |
| $E^1(d_0)$ | $\varphi$ | $\Phi$ | $\varphi$ |
| $E^1(\varepsilon)$ | $\varphi$ | $\Phi$ | $\{b\}$ |
| $E^1(a)$ | $\varphi$ | $\{a\}$ | $\{a\}$ |
| $E^1(b)$ | $\{\varepsilon\}$ | $\{\varepsilon\}$ | $\{\varepsilon, b\}$ |
| $E^1(aa)$ | $\{\varepsilon\}$ | $\{\varepsilon\}$ | $\{\varepsilon, b\}$ |
| $E^1(ab)$ | $\Phi$ | $\{a\}$ | $\{a\}$ |

Table 2.5.1(a'): For $B_1$

Table 2.5.1(a') is now consistent as it finds no such pair that is $E_i(\alpha) = E_i(\beta)$ but $E_i(f(\alpha, \sigma)) \mathrel{/}= E_i(f(\beta, \sigma))$. So, it constructs the hypothesis DFA $H_m$ by taking product of $B_1$, $B_2$, $B_3$. For this, the IKL algorithm constructs the 1 bit automata $B_1$, $B_2$, $B_3$ for each corresponding table which are given below in Figure 2.5.3.

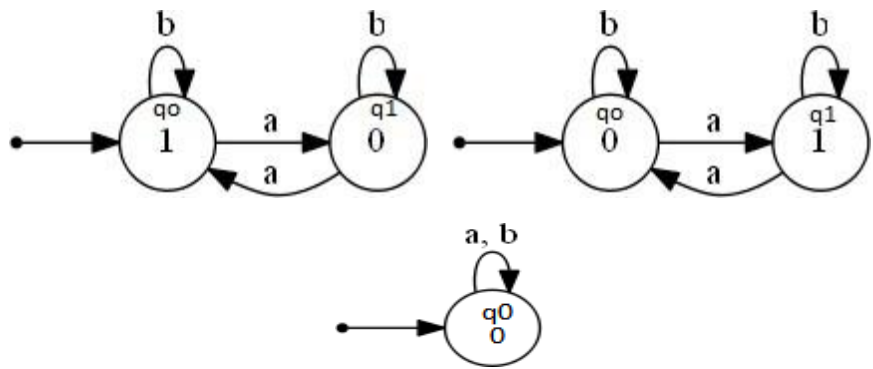

Figure 2.5.3: $B_1$, $B_2$, $B_3$

Product of all bit slice automata's $B_1'$, $B_2'$, $B_3'$ is given below in Figure 2.5.4.

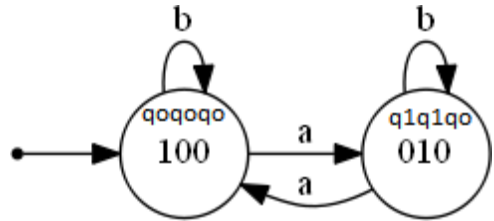

Figure 2.5.4: Product Automata $H_m$

We can see that product automaton $H_m$ is behaviorally equivalent to the target automata $A$.

Now suppose the IKL algorithm reads the input string $ba$ as it is consistent with $H_m$ so $H_{m+1} = H_m \equiv A$. As the file $S$ is now empty so the IKL algorithm stops its execution.

## 2.6 RPNI Algorithm

The RPNI algorithm is a passive learning algorithm proposed by Jose Oncina and Pedro Garcia in 1992 [3]. It uses a tree structure instead of table and does not maintain consistency. It takes the input as set of positive examples and set of negative examples $S_+$ and $S_-$ respectively. It first writes the elements of $S_+$ and its prefixes in lexicographical order then from set of positive examples and their prefixes, it constructs the prefix tree $PT(S_+)$. After that it recursively partitions the branches of the tree into blocks. The partition is represented as $\pi$ and the target automata is represented as $A$. At first step each element of $PT(S_+)$ belongs to its self-containing block. The RPNI algorithm recursively applies joint operation on these blocks so that they can be merged into two final blocks. One is accepting state block and second is non-accepting state block.

Let $\pi$ be a partition over $PT(S_+)$ and blocks $B_i, B_j \in \pi$ then joint operation over any two blocks $B_i, B_j$ is $J(\pi, B_i, B_j) = \{ B \in \pi \mid B \mathrel{/}= B_i, B \mathrel{/}= B_j \} \cup \{ B_i \cup B_j \}$. Initial automaton $A_0$ produced by $PT(S_+) = \pi_0 = \{ u_0, u_1, \ldots, u_r \}$ and $\pi_n = J(\pi_{n-1}, B, u_n)$ i $S_-$   $L(A_0 / J(\pi_{n-1}, B, u_n)) = \varphi$ otherwise $\pi_n = \pi_{n-1}$. Detailed explanation of the RPNI algorithm is given below with the help of example.

### Example

An example run of the RPNI algorithm is given below:
   Target Automaton $A$: Odd number of a's
   $S_+ = \{$ a, ab, bab, abaa $\}$
   $S_- = \{$ b, baba, baa $\}$

The lexicographical order of prefixes of $S_+$ is:
⟨ $\lambda$, a, b, ab, ba, bab, aba, abaa⟩
Initial automaton $A_0 = PT(S_+)$ is given in Figure 2.6.1.

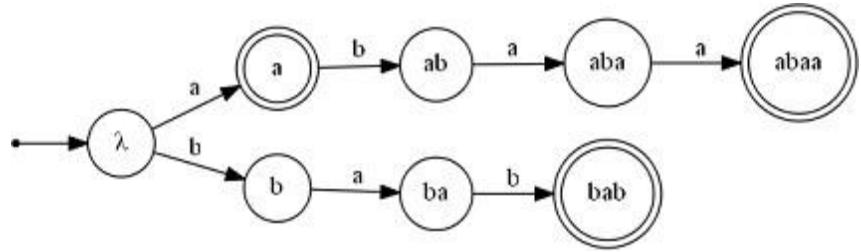

Figure 2.6.1: Initial Automata $A_0$ / $\pi_0$

To obtain $\pi_1$ where $u_1$= "a" , the RPNI algorithm perform operation $J$ ( $\pi_0, \lambda , a$ ) which is given in Figure 2.6.2.

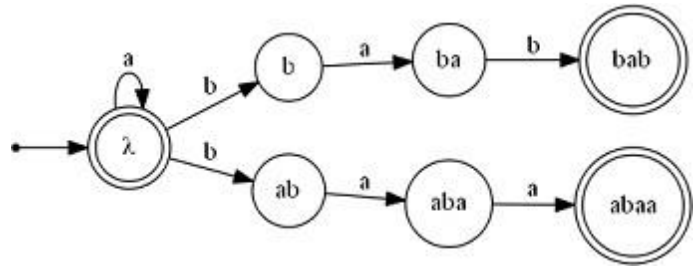

Figure 2.6.2: $A_0$ / $J$ ( $\pi_0, \lambda, a$ )

In Figure 2.6.2, we can see that $S_-$ ⊤ $L( A_0 / J ( \pi_0, \lambda, a))$ /= $\varphi$ as above automaton accepts the string *baa* which belongs to the set $S_-$. As there are no more states to try to merge with $u_1$= "a" therefore $\pi_1 = \pi_0$

To obtain $\pi_2$ where $u_2$= "b" , the RPNI algorithm performs operation $J$ ( $\pi_1, \lambda, b$ ) which is given in Figure 2.6.3.

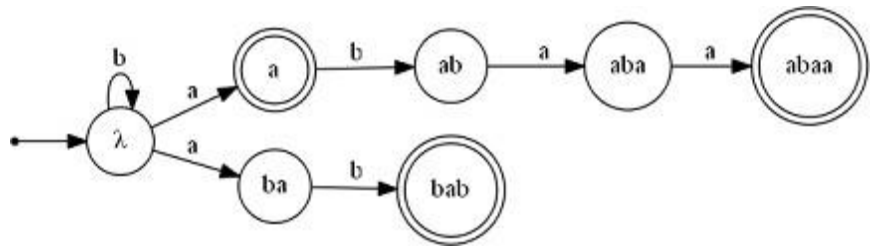

Figure 2.6.3: $A_0$ /$J$ ( $\pi_1, \lambda, b$ )

In Figure 2.6.3, we can see that $S_- \sqcap L(A_0 / J(\pi_1, \lambda, b)) = \varphi$ as above automata accepts all strings belonging to the set $S_+$ and rejects all negative data belonging to the set $S_-$. So $\pi_2 = J(\pi_1, \lambda, b)$.

To obtain $\pi_3$ where $u_3 = $ "ab", the RPNI algorithm performs operation $J(\pi_2, a, ab)$ which is given in Figure 2.6.4.

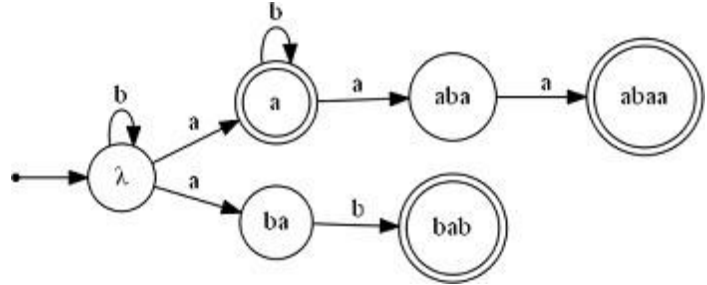

Figure 2.6.4: $A_0 / J(\pi_2, a, ab)$

In Figure 2.6.4, we can see that $S_- \sqcap L(A_0 / J(\pi_2, a, ab)) = \varphi$ as above automaton accepts all strings belonging to the set $S_+$ and rejects all negative data belonging to the set $S_-$. Therefore $\pi_3 = J(\pi_2, a, ab)$.

To obtain $\pi_4$ where $u_4 = $ "ba", the RPNI algorithm performs operation $J(\pi_3, a, ba)$ which is given in Figure 2.6.5.

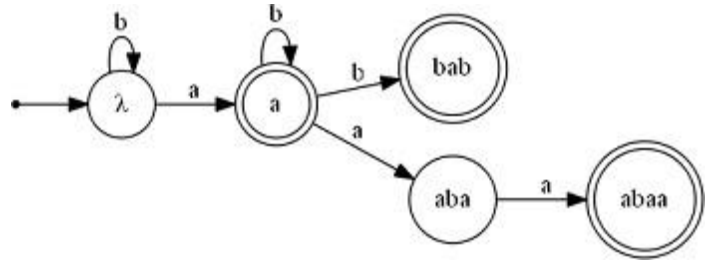

Figure 2.6.5: $A_0 / J(\pi_3, a, ba)$

In Figure 2.6.5, we can see that $S_- \sqcap L(A_0 / J(\pi_3, a, ba)) = \varphi$ as above automaton accepts all strings belonging to the set $S_+$ and rejects all negative data belonging to the set $S_-$. Therefore $\pi_4 = J(\pi_3, a, ba)$.

To obtain $\pi_5$ where $u_5 = $ "bab", the RPNI algorithm performs operation $J(\pi_4, a, bab)$ which is given in Figure 2.6.6.

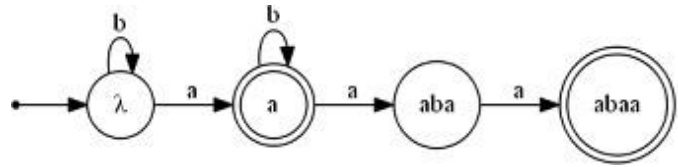

Figure 2.6.6: $A_0 / J(\pi_4, a, bab)$

In Figure 2.6.6, we can see that $S_- \top L(A_0 / J(\pi_4, a, bab)) = \varphi$ as above automaton accepts all strings belonging to the set $S_+$ and rejects all negative data belonging to the set $S_-$. Therefore $\pi_5 = J(\pi_4, a, bab)$.

To obtain $\pi_6$ where $u_6 =$ "*aba*", the RPNI algorithm performs operation $J(\pi_5, \lambda, aba)$ which is given in Figure 2.6.7.

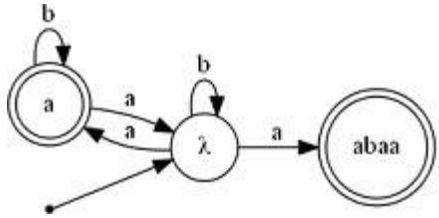

Figure 2.6.7 : $A_0 / J(\pi_5, \lambda, aba)$

In Figure 2.6.7, we can see that $S_- \top L(A_0 / J(\pi_5, \lambda, aba)) = \varphi$ as above automaton accepts all strings belonging to the set $S_+$ and rejects all negative data belonging to the set $S_-$. Therefore $\pi_6 = J(\pi_5, \lambda, aba)$.

To obtain $\pi_7$ where $u_7 =$ "*abaa*", the RPNI algorithm performs operation $J(\pi_6, a, abaa)$ which is given in Figure 2.6.8.

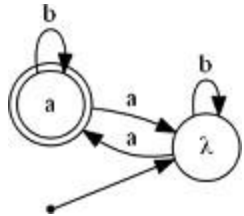

Figure 2.6.8 : $A_0 / J(\pi_6, a, abaa)$

In Figure 2.6.8, we can see that $S_- \top L(A_0 / J(\pi_6, \lambda, abaa)) = \varphi$ as above automaton accepts all strings belonging to the set $S_+$ and rejects all negative data belonging to the set $S_-$. Therefore $\pi_7 = J(\pi_6, a, abaa)$.

As there are total eight elements in lexicographical order in the pre xes of $S_+$ set so after seven recursive partitions $\pi_7$ the RPNI algorithm stops its execution and we can see that this partition is behaviorally equivalent to the target automaton $A$.

## 2.7 RPNII Algorithm

The RPNII algorithm is an incremental extension of the RPNI algorithm [3]. The RPNI algorithm takes the positive and negative examples as a whole and can't accommodate new labeled example unless it may start its whole execution from the scratch. The RPNII algorithm reduces this discrepancy as it has the ability to accommodate a new labeled example easily [4].

The RPNII algorithm initially takes the set of positive and negative examples, $S_+$, $S_-$ respectively. It also takes the prefix tree acceptor PTA($S_+$), deterministic quotient automaton (DQA) and a new labeled example x.

If the new labeled example consistent with deterministic quotient automaton (DQA) then initial deterministic quotient automaton will be the final solution. Otherwise, the RPNII algorithm accommodates new labeled example by recursive splitting process in form of depth first search (in reverse lexicographical order). This process continuous until the quotient automaton becomes deterministic. After that when the quotient automaton becomes deterministic as well as consistant with $S_+$ and $S_-$ then the RPNII algorithm applies the RPNI algorithm on it. Which we have brie y explained in the previous section.

### Example

An example run of the RPNII algorithm is described below:

Let $S_+$ = { $\lambda$, ab, bab, babb }

$S_-$ = { a, baa }

The lexico-graphical order = ⟨ $\lambda$, a, b, ab, ba, bab, babb ⟩

PTA($S_+$):

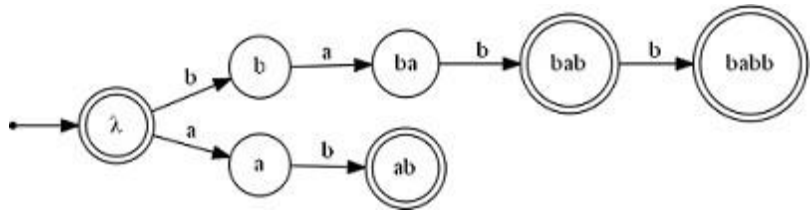

Figure 2.7.1: PTA($S_+$)

DQA:

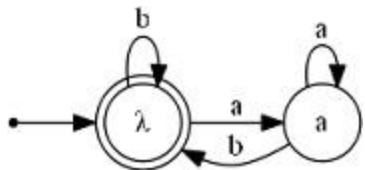

Figure 2.7.2: Initial Deterministic Quotient Automaton

let x = ( *b*, —)

As initial DQA shows that b is accepting string while new labeled example x shows that string b belongs to $S_-$ therefore the RPNII algorithm modifies the initial DQA to make it consistent with the sets $S_+$ and $S_-$. For this purpose, the RPNII algorithm starts from the string *babb* and splits the initial deterministic quotient automaton which is given below in Figure 2.7.3.

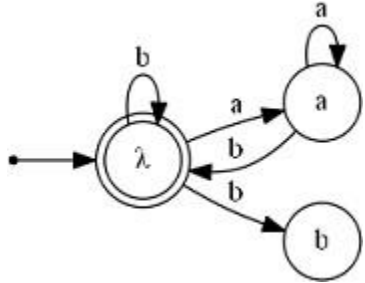

Figure 2.7.3: Splitting for the string *babb*

Figure 2.7.3 shows that due to splitting of initial DQA for the string *babb*, the initial automaton became non-deterministic as the initial state has two transitions for input symbol b. Therefore, to make it deterministic, the RPNII algorithm again splits this automaton on the basis of the string *bab* which is given below in Figure 2.7.4.

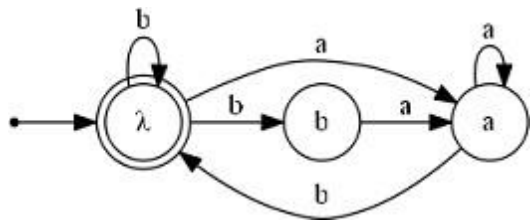

Figure 2.7.4: Splitting for the string *bab*

Figure 2.7.4 shows that splitting at the string *bab* also creates non-determination at the initial state, as this state has two transitions for input symbol b. Therefore, the RPNII algorithm again splits this automaton on the basis of the string *ba*. New quotient automaton is described below in Figure 2.7.5.

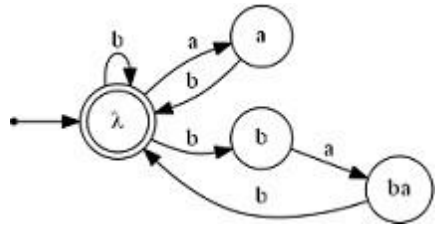

Figure 2.7.5: Splitting for the string *ba*

Figure 2.7.5 shows that splitting at the string *ba* also create non-determination at the initial state, as this state has two transitions for input symbol b . Therefore the RPNII algorithm again splits this automaton on the basis of the string *b*. New quotient automaton is described below in Figure 2.7.6.

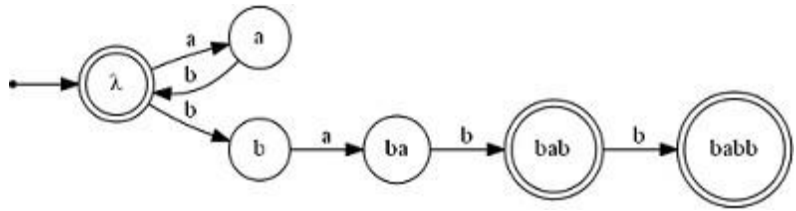

Figure 2.7.6: Splitting for the string *b*

As we can see that Figure 2.7.6 shows the quotient automaton $A_0$ which is now deterministic and consistent with the sets, $S_+$ and $S_-$. Therefore, the RPNII algorithm stops its recursive splitting process.

Here the RPNII algorithm applies the RPNI algorithm on deterministic quotient automaton which is given below.

The lexicographical order of automaton $A_0$ is $\langle \lambda, a, b, ba, bab, babb \rangle$

$u_1 = a$

$J(\pi_0, \lambda, a)$

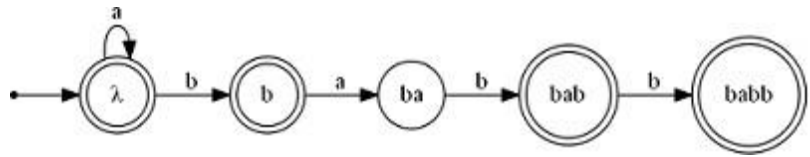

Figure 2.7.7: $A_0 / J(\pi_0, \lambda, a)$

Figure 2.7.7 shows that $L(A_0/J(\pi_0, \lambda, a)) \cap S_- \neq \varphi$ as strings $a$ and $b$ are accepting here, according to Figure 2.7.7 but these belong to $S_-$. Therefore $\pi_1 = \pi_0$

Now $u_2 = b$

$J(\pi_1, \lambda, b)$

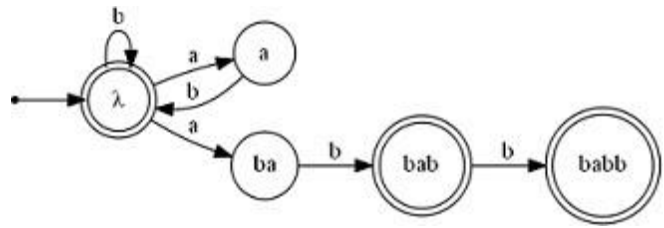

Figure 2.7.8: $A_0 / J(\pi_1, \lambda, b)$

Figure 2.7.8 shows that $L(A_0/J(\pi_1, \lambda, b)) \cap S_- \neq \varphi$ as string $b$ is accepting in Figure 2.7.8 but this is belonging to $S_-$. Therefore $\pi_2 = \pi_1$

Now $u_3 = ba$

$J(\pi_2, b, ba)$

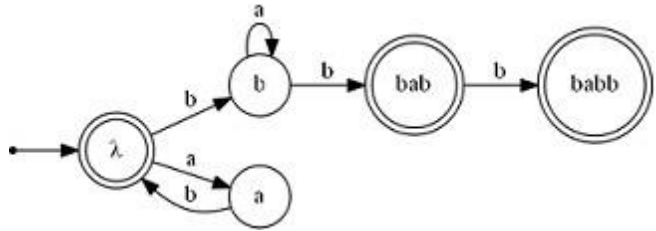

Figure 2.7.9: $A_0 / J(\pi_2, b, ba)$

Figure 2.7.9 shows that $L(A_0/J(\pi_2, b, ba)) \cap S_- = \varphi$ therefore we can say that it rejects all negative strings. So $\pi_3 = J(\pi_2, b, ba)$

Now $u_4 = bab$

$J(\pi_2, ba, bab)$

This operation is not suitable as if we will merge the strings $ba$ and $bab$ then in the next step, string $b$ will be accepted but as it belongs to $S_-$ so it should not be accepted.

Now $u_4 = babb$

$J(\pi_2, bab, babb)$

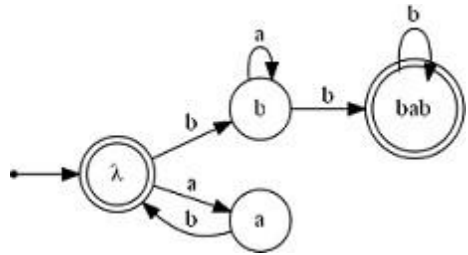

Figure 2.7.10: $A_0 / J(\pi_2, bab, babb)$

Figure 2.7.10 shows that this automaton $H = A_0/J(\pi_2, bab, babb)$ is behaviorally equivalent to the target automaton $A$ as well as it is also consistent with the $S_+$ and $S_-$ sets. Therefore, RPNII algorithm stops its execution and returns the automaton $H$ as an output.

## 2.8 Analysis

### 2.8.1 Time Complexities of Learning Algorithms

| Algorithm | Learning Type | Complexity | Learned Automata |
|---|---|---|---|
| L* | Complete | $O(|\Sigma|.N^2 M)$ | Moore |
| ID | Complete | $O(|\Sigma|.|P|.N)$ | Moore |
| IID | Incremental | $O(|\Sigma|.|P_l|.N)$ | Moore |
| IDS | Incremental | $O(|\Sigma|.|P_k|.N)$ | Moore |
| IKL | Incremental | $O(|\Sigma|.|P|.Nl)$ | Moore |
| RPNI | Complete | $O((|I_p|+|I_n|).|I_p|^2)$ | Moore |
| RPNII | Incremental | $O((|I_p|+|I_n|).|I_p|^2)$ | Moore |

Table 2.8.1: Complexities of Learning Algorithms

Above Table 2.8.1 shows that size of the input alphabet $|\Sigma|$, number of nodes $N$ in the target DFA $A$ and the number of queries; $M$ for L*, $|P|$ for the ID and IKL algorithms, $|P_l|$ for IID algorithm, $|P_k|$ for IDS algorithm, $|I_p|$ (pos- itive sample) and $|I_n|$ (negative sample) for the RPNI and RPNII algorithms, contribute in the complexities of learning algorithms. If we analyze, we can see that number of queries have major contribution in the complexities of above mentioned algorithms as the size of input alphabet $|\Sigma|$ and number of nodes $N$ in the target DFA are nearly static factors.

### 2.8.2 Query-Wise Analysis of Learning Algorithms

| Algorithm | Membership Queries | Book-keeping Queries | Lexicographical Order |
|---|---|---|---|
| L* | Yes | No | No |
| ID | Yes | No | No |
| IID | Yes | Yes | No |
| IDS | Yes | Yes | No |
| IKL | Yes | Yes | No |
| RPNI | No | No | Yes |
| RPNII | No | No | Yes |

Table 2.8.2: Query wise Analysis of Learning Algorithms